\documentclass[preprint,prl,showpacs]{revtex4}

\usepackage{graphicx}
\usepackage{amsmath,amssymb}
\usepackage{color}
\usepackage{bm}

\renewcommand{\div}{\nabla \cdot }
\newcommand{\grad}{\nabla}
  
\newcommand{\Br}{{\bf x}}

\newcommand{\Bk}{{\bf k}}

\newcommand{\BA}{{\bf A}}

\begin{document}

\title{Ultrasound Modulated Bioluminescence Tomography}

\author{Guillaume Bal}
\affiliation{Department of Applied Physics and Applied Mathematics, Columbia University, New York, NY 10027, USA}
\email{gb2030@columbia.edu}

\author{John C. Schotland}
\affiliation{Departments of Mathematics and Physics, University of Michigan, Ann Arbor, MI 48109, USA }

\email{schotland@umich.edu}

\date{\today}

\begin{abstract}
We propose a method to reconstruct the density of a luminescent source in a highly-scattering medium from ultrasound modulated optical measurements. Our approach is based on the solution to a hybrid inverse source problem for the diffusion equation.
\end{abstract}

\pacs{42.30.Wb}

\maketitle

The development of tools for molecular imaging has had a transformative effect on biomedical research~\cite{weissleder_2001}. There are multiple applications including mapping gene expression and following the course of infection in a single animal, among others. Optical methods hold great promise for molecular imaging, due to their spectroscopic sensitivity to chemical composition, nondestructive nature and relatively low cost~\cite{ntziachristos_2005}. One particularly popular modality, known as bioluminescence imaging, makes use of a bioluminescent marker, most often the luciferin-luciferase system, as a reporter of molecular activity~\cite{contag_2002,mccafrey_2003}. In a typical experiment, genetically modified light-emitting cells are introduced into a model organism and a CCD camera is used to record the intensity of emitted light. The resulting images convey information about the spatial distribution of the labeled cells. However, the images are not tomographic nor are they quantitatively related to the number density of the cells. One approach to this problem is to reconstruct the number density (optical source) from measurements of multiply-scattered light, a method known as bioluminescence tomography (BLT)~\cite{wang_2003,wang_2004,gu_2004,congl_2005,jiang_2007,ahn_2008,dehghani_2008,lu_2009,shi_2013}. The corresponding inverse problem is a classical inverse source problem (ISP) and it is well known that such problems do not have unique solutions~\cite{isakov_book}. That is, more than one source can give rise to the same measurements. Uniqueness can be restored under strong mathematical assumptions requiring \emph{a priori} knowledge of the source geometry. 

To overcome the problem of non uniqueness in BLT requires a fundamentally new approach. In this Letter, we propose a novel imaging modality termed ultrasound modulated bioluminescence tomography (UMBLT), which is in the spirt of several recently developed hybrid imaging methods. In hybrid imaging (also called multi-wave imaging), an external field is used to control the material properties of a medium of interest, which is then probed by a second field~\cite{wang_book,ammari_2008,capdeboscq_2009,bal_2010,bal_2012_a,bal_2010,bal_2012_b,bal_2012_c,bal_2013,cox_2009,gebauer_2009,kuchment,kuchment_2012,monard_2011,mclaughlin_2004,mclaughlin_2010,nachman,moskow_2013}. In the physical setting we consider, the source density is spatially modulated by an acoustic wave, while measurements of the emitted light are recorded. We find that it is possible to \emph{uniquely} reconstruct the source density by an algebraic formula. Moreover, the reconstruction is stable in the sense that an error in the measurements is linearly related to the error in recovering the source. 

We note that our results are particularly timely in view of recent exciting work by Huynh et al.~\cite{huynh_2013}. These authors report experiments in which a focused ultrasound beam is used to enhance the resolution of bioluminescence images. Such experiments provide the necessary input data to reconstruct the source density in UMBLT.

We begin by recalling the mathematical formulation of BLT. We consider a highly-scattering medium in which light propagates as a diffuse wave~\cite{arridge_2009}. The energy density $u$ of the wave is assumed to obey the time-independent diffusion equation
\begin{eqnarray}
\label{diff_eq}
-\div \left[ Dn^2 \grad\left( \frac{u}{n^2}\right)\right] + \alpha u  &=& S  \quad {\rm in} \quad \Omega \ , \\
u + \ell \frac{\partial u}{\partial n} &=& 0 \quad {\rm on} \quad \partial\Omega \ .
\label{bc}
\end{eqnarray}
Here $\Omega$ is a three-dimensional bounded domain, $n$ is the index of refraction,  $\alpha$ and $D$ are the absorption and diffusion coefficients of the medium, $S$ is the source density and $\ell$ is the extrapolation length.  We note that in bioluminescence imaging the source is \emph{incoherent} and emits light over a broad range of frequencies. Thus, for the remainder of this Letter, we assume that the intensity is measured over a relatively narrow band of frequencies so that the frequency-dependence of the absorption and diffusion coefficients can be neglected.

The inverse problem of BLT is to determine the source density $S$ everywhere in the volume $\Omega$ from measurements of the intensity on $\partial\Omega$. As previously mentioned, this problem does not have a unique solution, due to the existence of nonradiating sources; such sources generate fields that vanish everywhere in their exterior.  This difficulty may be overcome, to some extent, if it is known that $S$ is constant on a fixed number of regions of known shape. It is also possible to determine geometrical properties of the source, such as its spatial extent.  

To address the above mentioned difficulties, we introduce an acoustic wave field that spatially modulates the source. This \emph{internal control} of the medium provides information that is not available in conventional ISPs. 
To proceed, we consider the medium to be a collection of particles (cells) suspended in a fluid in which the acoustic wave propagates. Some of the particles absorb and scatter light, while others act as sources and emit light.  If a small amplitude acoustic wave is incident on the medium, then each particle will experience an acoustic radiation force and oscillate about its local equilibrium position. We assume that the acoustic pressure is a standing plane wave of the form
$p=A\cos(\omega t)\cos(\Bk\cdot \Br +\varphi)$, where $\omega$ is the frequency, $A$ is the amplitude, $\Bk$ is the wavevector and $\varphi$ is the phase of the wave. For simplicity, we have assumed that the speed of sound $c_s$ is constant with $k=\omega/c_s$. 
If the particles have positions $\Br_i$, then their number density is 
$\rho(\Br)=\sum_i \delta(\Br-\Br_i)$.
It can be seen that the number density is spatially modulated
according to
\begin{equation}
\label{rho}
\rho_\epsilon(\Br)=\rho_0(\Br)\left[1+ \epsilon \cos(\Bk\cdot \Br +\varphi)\right] \ ,
\end{equation}
where $\rho_0$ is the number density in the absence of the acoustic wave and $\epsilon=A/(\rho c_s^2) \ll 1$ is a small parameter~\cite{bal_2010}. Now, the source density is proportional to the density of light-emitting cells and is thus given by
\begin{equation}
S_\epsilon(\Br) = S_0(\Br)\left[1+ \epsilon \cos(\Bk\cdot \Br +\varphi)\right] \ ,
\end{equation}
where $S_0$ is the source density in the absence of the acoustic wave. The optical properties of the medium are also acoustically modulated. In particular, the index of refraction of the fluid in which the particles are suspended is modulated due to Brillouin scattering and is given by
\begin{equation}
\label{index}
n(\Br) = n_0\left[1+ \epsilon \gamma\cos(\Bk\cdot \Br +\varphi)\right] \ ,
\end{equation}
where $n_0$ is the unmodulated index of refraction and $\gamma$ is the elasto-optical constant.
We note that $\gamma \approx 0.3$ in water.  
In~\cite{bal_2010} it was shown that the absorption and diffusion coefficients are modulated according to
\begin{eqnarray}
\alpha_\epsilon(\Br) &=&   \alpha_0(\Br)\left[1 + \epsilon(2\gamma + 1) \cos(\Bk\cdot \Br +\varphi)\right]    \ , \\
D_\epsilon(\Br) &=&   D_0(\Br)\left[1 + \epsilon(2\gamma - 1) \cos(\Bk\cdot \Br +\varphi)\right]  \ .
\end{eqnarray}
Making use of the above results, we see that (\ref{diff_eq}) and (\ref{bc}) become
\begin{eqnarray}
\label{diff_eqn_mod}
-\div D_\epsilon \grad u_\epsilon + \alpha_\epsilon u_\epsilon &=& S_\epsilon \quad {\rm in} \quad \Omega \ , \\
u_\epsilon + \ell \frac{\partial u_\epsilon}{\partial n} &=& 0 \quad {\rm on} \quad \partial\Omega \ ,
\label{bc_mod}
\end{eqnarray}
where $u_\epsilon = u/n^2$.

The inverse problem  is to recover $S_0$ from knowledge of $u_\epsilon$ on $\partial\Omega$. 
Here we assume that $\alpha_0$ and $D_0$ are known everywhere in $\Omega$ as determined, for instance, by an optical tomography experiment. 
It will prove useful to consider the auxiliary problem 
\begin{eqnarray}
\label{diff_eqn_aux}
-\div D_0 \grad v_j + \alpha_0 v_j &=& 0  \quad {\rm in} \quad \Omega \ , \\
v_j + \ell  \frac{\partial v_j}{\partial n} &=& f_j \quad {\rm on} \quad \partial\Omega \ , \quad j=1,\ldots,N \  ,
\label{bc_aux}
\end{eqnarray}
where $f_j$ are boundary sources. If we multiply (\ref{diff_eqn_aux}) by $u_\epsilon$ and (\ref{diff_eqn_mod}) by $v_j$, take the difference of the resulting equations and integrate over $\Omega$, we obtain the identity
\begin{equation}
\Sigma_\epsilon^{(j)} = \int_\Omega d^3x \left[\left(D_\epsilon-D_0\right ) \grad u_\epsilon\cdot\grad v_j  +\left(\alpha_\epsilon - \alpha_0\right)u_\epsilon v_j - v_j S_\epsilon \right] \ ,
\end{equation}
where we have integrated by parts and applied the boundary conditions (\ref{bc_mod}) and (\ref{bc_aux}). 
The surface term $\Sigma_\epsilon^{(j)}$ is defined by
\begin{equation}
\label{def_Sigma}
\Sigma_\epsilon^{(j)} = \int_{\partial\Omega} d^2x \left[  
u_\epsilon D_0  \frac{\partial v_j }{\partial n}-v_j D_\epsilon  \frac{\partial u_\epsilon}{\partial n}\right] \ .
\end{equation}
Next, we perform an asymptotic expansion of $u_\epsilon$ and $\Sigma_\epsilon^{(j)}$ in the small parameter $\epsilon$: 
\begin{eqnarray}
u_\epsilon &=& u_0 + \epsilon u_1 + \epsilon^2 u_2 + \cdots \ , \\
\Sigma_\epsilon^{(j)} &=& \Sigma_0^{(j)} + \epsilon \Sigma_1^{(j)} + \epsilon^2 \Sigma_2^{(j)} +  \cdots \  .
\end{eqnarray}
We find that to order $O(1)$ 
\begin{equation}
\Sigma_0^{(j)} = \int_\Omega d^3x v_j S_0   \ .
\end{equation}
At $O(\epsilon)$ we have
\begin{equation}
\Sigma_1^{(j)}(\Bk) = \int_\Omega d^3x \left[(2\gamma -1) D_0 \grad u_0\cdot\grad v_j + (2\gamma +1) \alpha_0 u_0 v_j - v_j S_0\right]
\cos\left(\Bk\cdot\Br + \varphi\right) \ .
\end{equation}

The intensity measured by a point detector on $\partial\Omega$, which collects light in the outward normal direction, is given by $I_\epsilon=c/(4\pi)(1+\ell^*/\ell)u_\epsilon$~\cite{markel_2004}. Here $\ell^*$ is the transport length, which is related to the diffusion coefficient by $D=1/3c\ell^*$. Making use of the boundary conditions (\ref{bc}) and (\ref{bc_aux}) we see that (\ref{def_Sigma}) becomes
\begin{equation}
\Sigma_\epsilon^{(j)} = \frac{4\pi}{3}\frac{\ell^*}{\ell+\ell^*}\int_{\partial\Omega}d^2x f_j I_\epsilon\cos(\Bk\cdot\Br + \varphi) \ .
\end{equation}
Evidently $\Sigma_1^{(j)}$ can be determined from experiment. Thus, by varying
the wave vector $\Bk$ and the phase $\varphi$ and inverting a Fourier transform, we can recover the so-called internal functional
\begin{equation}
\label{internal_functional}
H_j = (2\gamma-1)D_0 \grad u_0 \cdot\grad v_j +(2\gamma+1) \alpha_0 u_0 v_j - v_j S_0 
\end{equation}
from measurements. That is,
\begin{equation}
H_j(\Br) = \int \frac{d^3k}{(2\pi)^3}e^{-i\Bk\cdot\Br}\left[\Sigma_1^{(j)}(\Bk;0) + i\Sigma_1^{(j)}(\Bk;3\pi/2)\right]\ ,
\end{equation}
where the dependence of $\Sigma_1^{(j)}$ on $\varphi$ has been made explicit. 

The inverse problem now consists of recovering the source 
$S_0$ from the internal functional $H_j$. We emphasize that this is an unusual inverse problem, since the data $H_j$ is known everywhere in $\Omega$. This situation can be compared with that of the ISP, where the data is known only on $\partial\Omega$. The ISP is thus underdetermined, which leads to the previously mentioned problem of non uniqueness. In contrast, we will see that the availability of internal data in UMBLT allows for the unique recovery of $S_0$.
\begin{figure}[t]
\vspace{-0.5in}
\centering
\includegraphics[width=3.0in]{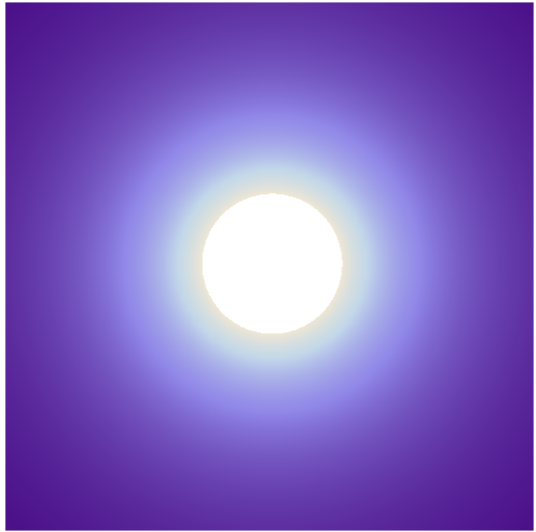}
\caption{(Color online) Reconstructed image of a point source. The field of view is $L/10\times L/10$.}
\label{fig3}
\end{figure}
We first consider the case of a single boundary source. Eq.~(\ref{internal_functional}) then becomes 
\begin{equation}
\label{def_H}
\frac{H}{v} = (2\gamma-1) D_0 \nabla\ln v \cdot \nabla u_0 + (2\gamma+1) \alpha_0 u_0 - S_0 \ ,
\end{equation}
which is well defined since $v$ does not vanish in $\Omega$.
Using the fact that
\begin{equation}
\label{S_0}
S_0 = -\div D_0\grad u_0 + \alpha_0 u_0  \ ,
\end{equation}
we can eliminate $S_0$ from (\ref{def_H}). We then find that $u_0$ obeys
the equation
\begin{eqnarray}
\label{PDE_L}
-(L-2\gamma\alpha_0)u_0 &=& \frac{H}{v} \quad {\rm in} \quad \Omega \ , \\
u_0 + \ell \frac{\partial u_0}{\partial n} &=& 0 \quad {\rm on} \quad \partial\Omega \ ,
\end{eqnarray}
where $Lu_0 := -\div D_0 \grad u_0 -(2\gamma-1)D_0\grad\ln v\cdot\grad u_0$. If $0$ is not an eigenvalue of $L-2\gamma\alpha_0$ with the above prescribed boundary conditions (which holds with suitable smallness conditions on $\alpha_0$ or $\Omega$)~\cite{supplement}, we can uniquely solve (\ref{PDE_L}) for $u_0$ with
\begin{equation}
u_0 = - (L-2\gamma\alpha_0)^{-1} \frac{H}{v} \ .
\end{equation} 
Once $u_0$ is known, we can obtain the source $S_0$ from
(\ref{S_0}). It follows immediately that $S_0$ can be reconstructed
with Lipschitz stability. That is, errors in $H$ propagate linearly to
errors in $S_0$. More precisely, suppose that $H$ and $H'$ are the internal data corresponding to the sources $S_0$ and $S_0'$, respectively. We then have the stability estimate
\begin{equation}
\label{stability}
\|S_0-S_0'\|_{L^2(\Omega)} \leq C \|H- H' \|_{L^2(\Omega)} \ ,
\end{equation} 
where $C$ is a fixed constant~\cite{supplement}.
See~\cite{supplement} for the case when $)$ is an eigenvalue of $L-2\gamma\alpha_0$.

Next we consider the inverse problem with multiple boundary sources. 
Note that since the coefficients $\alpha_0$ and $D_0$ are assumed to be known, the solutions $v_j$ can be computed numerically and thus additional experiments do not need to be performed.
To proceed, we assume that $(\grad v_j,v_j)$ form a basis for every point in $\Omega$.
It can be seen that this condition holds if the boundary sources $f_j$ are appropriately chosen~\cite{supplement}.  Assuming this is the case, (\ref{internal_functional}) forms a system of linear equations 
for the vector field $\BA=(2\gamma-1)D_0\grad u_0$ and the function $f=(2\gamma+1)\alpha_0u_0 - S_0$ of the form $Mg = H$. Here
$g=(f,\BA)$, $H=(H_1,\ldots,H_4)$ and
\begin{equation}
M =   
\left(
\begin{array}{ccc}
v_1 &  & (\grad v_1)^t \\ 
&  \ \vdots &  \\
v_4 &  & (\grad v_4)^t
\end{array}
\right) \ .
\end{equation}
Solving the above equations for $f$ and $\BA$ we obtain
\begin{eqnarray}
f &=& \sum_j (M^{-1})_{1j}H_j \ , \\
A_i &=& \sum_j (M^{-1})_{i+1,j}H_j  \ .
\end{eqnarray}
Since $\BA/D_0$ is a gradient field, it follows that
\begin{equation}
\label{integration}
u_0(\Br)-u_0(\Br_0) = \frac{1}{2\gamma-1 } \int_\Gamma \frac{1}{D_0} \BA \cdot d\Br \ ,
\end{equation}
where $\Gamma$ is an arbitrary path beginning at a point $\Br_0\in\Omega$ and ending at $\Br$. 
Using the above results, we find that the source $S_0$ may be obtained from the formula 
\begin{equation}
S_0(\Br) = S_0(\Br_0) + (2\gamma+1)\left[\alpha_0(\Br)u_0(\Br)- \alpha_0(\Br_0)u_0(\Br_0)\right]
-f(\Br)+f(\Br_0) \ ,
\end{equation}
which is the main result of this Letter. As before, it is readily seen that $S_0$ can be reconstructed with Lipschitz stability. 
The corresponding stability estimate is of the form
\begin{equation}
\label{stability_estimate}
\|S_0-S_0'\|_{L^2(\Omega)} \leq C \sum_j \|H_j- H_j'\|_{L^2(\Omega)}\ ,
\end{equation} 
where we have assumed that $S_0(\Br_0)=S_0'(\Br_0)$.

\begin{figure}[t]
\centering
\includegraphics[width=3in]{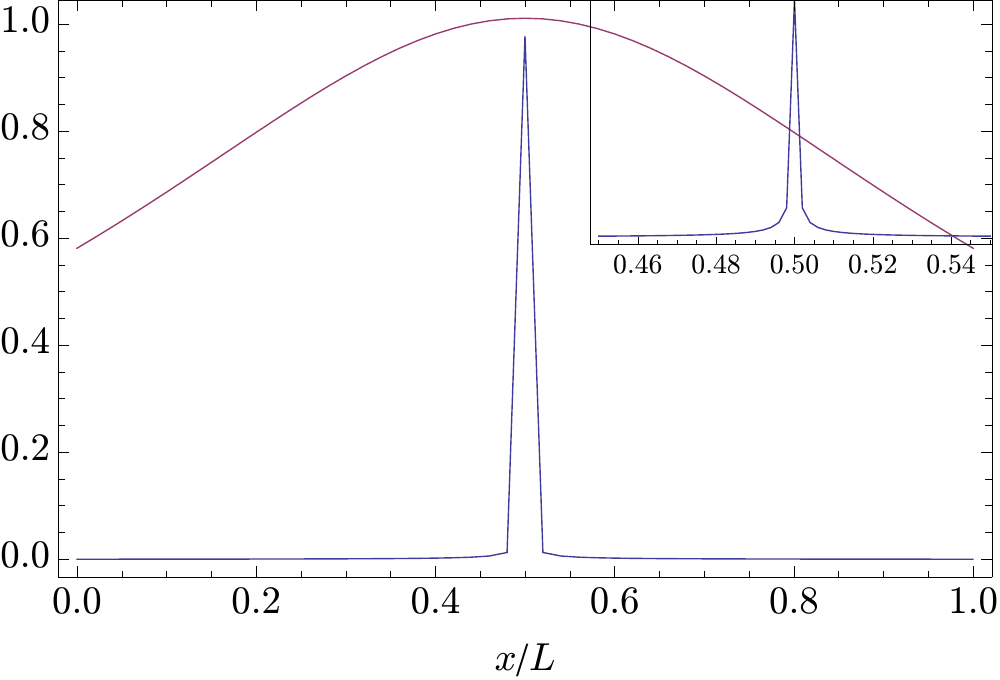} 
\caption{(Color online) One-dimensional profile of the reconstructed source in arbitrary units. The inset shows the peak in greater detail. The curve shown in red is the transmitted intensity due to the source. }
\label{fig5}
\end{figure}

We now illustrate the above reconstruction procedure with numerical simulations. For simplicity, we consider the case of an infinite homogeneous medium. The absorption and diffusion coefficients are
given by $\alpha_0 = 1.0 \ {\rm ns}^{-1}$ and $D_0=1.0 \ {\rm cm}^2 {\rm ns}^{-1}$, which is typical for biological tissue at optical wavelengths.
The $f_j$ are taken to be unit-amplitude point sources 
which occupy the vertices of a square of dimensions $L\times L$. 
Since the inverse problem is linear, it suffices to restrict our attention to a point source, which we place at the center of the square. In this setting, it is possible to compute the data $H_j$ 
in closed form. In Figure~1 we present a reconstruction of $S_0$ in
the plane containing the source. Here we take $L=1 \ {\rm cm}$ and the integration in (\ref{integration}) is performed with a step size
of $L/100$. Figure~2 shows a one-dimensional profile of the reconstructed source along a line passing through the center of the source. It can be seen that the resolution, as measured by the full width at half maximum (FWHM) is approximately $L/50$.
We note that this must be considered to be a best-case estimate since the effects of noise have not been considered. However, the stability estimate (\ref{stability_estimate}) indicates that there will be relatively little degradation of the resolution in the presence of noise. It is instructive to contrast the above results with those that can be obtained by conventional bioluminescence imaging. To this end, also shown in Figure~2 is the transmitted intensity due to the source measured on a line coinciding with an edge of the square region in which the measurements are performed. The FWHM of the intensity is
approximately $L/5$. Thus the resolution of the reconstructed image
is a factor of ten higher than in conventional bioluminescence imaging.

We close with several remarks. (i) In general, the absorption coefficient $\alpha_0$ and diffusion coefficient $D_0$ will not be known with high
spatial resolution, as would be the case if they were determined from optical tomography experiments~\cite{arridge_2009}. Thus, it would be of interest to determine the effect of errors in $\alpha_0$ and $D_0$ on reconstruction of the source $S_0$. (ii) The diffusion equation (\ref{diff_eq}) is valid when the energy density varies slowly on the scale of the transport mean free path. This condition breaks down when the acoustic wavelength is sufficiently small. It would thus be useful to extend the theory we have developed to the regime in which light propagation is described by the radiative transport equation~\cite{arridge_2009}.
(iii) In many biomedical applications, the speed of sound in tissue is not constant. Our results generalize straightforwardly to this case. In particular, we note that for known, sufficiently localized fluctuations in the sound speed, recovery of the internal functional is possible by a suitably modified Fourier transform~\cite{taylor_book}.

In conclusion, we have developed a hybrid imaging method for reconstructing the source density in bioluminescence tomography. Our approach is based on the solution to an inverse problem for the diffusion equation with interior control of boundary measurements.

G.B. was supported by NSF Grant DMS-1108608.  J.C.S. was supported by NSF grants DMS-1115574 and DMS-1108969.


\begin{thebibliography}{99}

\bibitem{weissleder_2001}
R. Weissleder and U. Mahmood, Radiology {\bf 219}, 316-333 (2001).

\bibitem{ntziachristos_2005}
V. Ntziachristos, J. Ripoll, L. H. V. Wang, and R. Weissleder, Nat. Biotech. {\bf 23}, 313-320 (2005).

\bibitem{contag_2002}
C. Contag and M. H. Bachmann, Annu. Rev. Biomed. Eng. {\bf 4}, 235-260 (2002).

\bibitem{mccafrey_2003}
A. McCaffrey, M. A. Kay and C. H. Contag, Molecuar Imaging {\bf 2}, 75-86 (2003).

\bibitem{wang_2003}
G. Wang, E. A. Hoffman, G. McLennan, L. V. Wang, M. Suter and J. Meinel, Radiology {\bf 229(P)}, 566 (2003).

\bibitem{wang_2004}
G. Wang, Y. Li and M. Jiang, Med. Phys. {\bf 31}, 2289-2299 (2004).

\bibitem{gu_2004}
X. Gu, Q. Zhang, L. Larcom and H. Jiang, Opt. Express {\bf 12}, 3996-4000 (2004).

\bibitem{congl_2005}
W. Cong1 et al. Optics Express {\bf 13}, 6756-6771 (2005).

\bibitem{jiang_2007}
Ming Jiang, Tie Zhou, Jiantao Cheng, Wenxiang Cong and Ge Wang, Optics Express {\bf 15}
11095-11116 (2007).

\bibitem{ahn_2008}
S. Ahn, A.J. Chaudhari, F. Darvas, C.A. Bouman and R.M. Leahy, Phys. Med. Biol. {\bf 53}, 3921-3942 (2008).

\bibitem{lu_2009}
Y. Lu, X. Zhang, A. Douraghy, D. Stout, J. Tian, T. F. Chan and A. F. Chatziioannou, Opt. Express {\bf 17}, 8062-8080 (2009).

\bibitem{dehghani_2008}
H. Dehghani, S. C. Davis and B. W. Pogue, Med. Phys. {\bf 35}, 4863-4871 (2008).

\bibitem{shi_2013}
S. Shi and H. Mao, Biomedical Optics Express {\bf 4}, 709-724 (2013).

\bibitem{isakov_book}
V. Isakov, Inverse Source Problems (American Mathematical Society, Providence, 1990).

\bibitem{wang_book}
L. H. Wang (Editor), Photoacoustic imaging and spectroscopy (CRC Press, 2009).

\bibitem{ammari_2008}
H. Ammari, E. Bonnetier, Y. Capdeboscq, M. Tanter and M. Fink, SIAM J. Appl. Math. {\bf 68}, 1557-1573 (2008).

\bibitem{capdeboscq_2009}
Y. Capdeboscq, J. Fehrenbach, F. de Gournay and O. Kavian, SIAM J. Imaging Sciences, {\bf 2}, 1003-1030 (2009).

\bibitem{bal_2010}
G. Bal and J. C. Schotland, Phys. Rev. Lett. {\bf 104}, 043902 (2010).

\bibitem{bal_2012_a}
G.~Bal in Inside Out II, G. Uhlmann Editor (Cambridge University Press,
Cambridge, UK, 2012).

\bibitem{bal_2010}
G. Bal, G. Uhlmann, Inverse Problems {\bf 26}, 085010 (2010).

\bibitem{bal_2012_b}
G.~Bal and G.~Uhlmann, Comm. Pure Appl. Math. {\bf 66}, 1629-1652 (2013).

\bibitem{bal_2013_b}
G.~Bal, Contemporary Mathematics (in press).

\bibitem{bal_2012_c}
G. Bal, E. Bonnetier, F. Monard and F. Triki, Inverse Problems and Imaging {\bf 7}, 353-375 (2013). 

\bibitem{bal_2013}
G. Bal, W. Naetar, O. Scherzer and J. Schotland, J. Ill-Posed and Inverse Problems {\bf 21}, 265280 (2013). 

\bibitem{cox_2009}
B. T. Cox, S. R. Arridge and P. C. Beard, J. Opt. Soc. Am. A, {\bf 26}, 443-455 (2009). 

\bibitem{gebauer_2009}
B. Gebauer and O. Scherzer, SIAM J. Applied Math. {\bf 69}, 565-576 (2009).

\bibitem{kuchment}
P. Kuchment and L. Kunyansky, J. Appl. Math. {\bf 19}, 191-224 (2008); ibid 
Inverse Problems {\bf 27} 055013 (2011).

\bibitem{kuchment_2012}
P. Kuchment and D. Steinhauer, Inverse Problems {\bf 28}, 084007 (2012). 

\bibitem{monard_2011}
F. Monard and G. Bal, Inverse Problems and Imaging {\bf 6}, 289-313 (2012).

\bibitem{mclaughlin_2004}
J. R. McLaughlin and J. Yoon, Inverse Problems {\bf 20}, 2545 (2004).

\bibitem{mclaughlin_2010}
J. R. McLaughlin, N. Zhang and A. Manduca, Inverse Problems {\bf 26}, 085007 (2010).

\bibitem{nachman}
Adrian Nachman, Alexandru Tamasan and Alexandre Timonov, Inverse Problems {\bf 23}, 2551-2563 (2007); 
ibid, Inverse Problems {\bf 25}, 035014 (2009).

\bibitem{moskow_2013}
J. C. Schotland and S. Moskow, Contemporary Mathematics (in press).

\bibitem{huynh_2013}
N. T. Huynh, B. R. Hayes-Gill, F. Zhang and S. P. Morgan, 
J. Biomedical Optics {\bf 18}, 020505 (2013).

\bibitem{markel_2004}
V. Markel and J. C. Schotland, Phys. Rev. E {\bf 70}, 056616
(2004).

\bibitem{arridge_2009}
S. R. Arridge and J. C. Schotland, Inverse Problems {\bf 25}, 123010 (2009).

\bibitem{supplement}
See the supplementary information.

\bibitem{taylor_book}
M. E. Taylor, Partial Differential Equations II: Qualitative
Studies of Linear Equations (Springer, New York, 1997), Chap. 9.

\end{thebibliography}
\end{document}


\title{Supplementary Information: \\ Ultrasound Modulated Bioluminescence Tomography}

\author{Guillaume Bal}
\affiliation{Department of Applied Physics and Applied Mathematics, Columbia University, New York, NY 10027, USA}

\author{John C. Schotland}
\affiliation{Departments of Mathematics and Physics, University of Michigan, Ann Arbor, MI 48109, USA }

\date{\today}
\maketitle

\newpage

\section{Solvability of (23)}

When $0$ is an eigenvalue of $L-2\gamma\alpha_0$, then (\ref{PDE_L}) cannot be solved uniquely. If $u_0$ and its normal derivatives can be measured on $\partial \Omega$, then we can alternatively solve the problem (which we can prove by unique continuation admits a unique solution)
\begin{eqnarray}
&&(L-2\gamma\alpha_0)^* (L-2\gamma\alpha_0) u_0 = -(L-2\gamma\alpha_0)^* \frac Hv \quad {\rm in } \quad \Omega \ , \\
&&u_0=g_0  \ , \quad \frac{\partial u_0}{\partial n}=j_0  \quad {\rm on } \quad\partial\Omega \ .
\end{eqnarray}
Once $u_0$ is known, then we can calculate $S_0$ as before. Standard elliptic estimates (see e.g. \cite{bal_2013_b}) combined with \eqref{eq:sfromu} show that errors in $H$, $g_0$ and $j_0$ propagate to errors in $S_0$ as
\begin{equation}
  \|S_0-S_0'\|_{L^2(\Omega)} \leq C \Big(\|H-H' \|_{L^2(\Omega)} + \|g_0- g_0' \|_{H^{\frac32}(\partial\Omega)} + \|j_0- j_0'\|_{H^{\frac12}(\partial\Omega)}\Big) \ ,
\end{equation} 
for a fixed constant $C$.

\section{Proof of the stability estimate (26)}

The estimate can be derived as follows:
\begin{eqnarray}\label{eq:sfromu}
\nonumber
&&\|S_0-S_0' \|_{L^2(\Omega)} \leq \|\alpha_0(u_0-u_0')\|_{L^2(\Omega)} +
\|\div D_0 \grad (u_0 -u_0') \|_{L^2(\Omega)} \\
\nonumber
&&\leq \|\alpha_0\|_{L^\infty(\Omega)}\|u_0-u_0'\|_{L^2(\Omega)} 
+  \|\grad D_0\|_{L^\infty(\Omega)}\|\grad (u_0-u_0')\|_{L^2(\Omega)} + \|D_0\|_{L^\infty(\Omega)}\|\lap(u_0-u_0')\|_{L^2(\Omega)}
 \\
&&\leq C \|u_0-u_0'\|_{H^2(\Omega)}  \ ,
\end{eqnarray}
for a suitable constant $C$. Since $L-2\gamma\alpha_0$ is smoothing by two orders, we obtain the required result.

\section{Proof that  $(\nabla v_j,v_j)$ forms a basis}

In the case that $D_0$ and $\alpha_0$ are constant, then $v$ is a solution of $-\Delta v + \frac{\alpha_0}{ D_0} v=0$. Let is introduce the quantity $\beta=\sqrt{\alpha_0/D_0}$. Then we find that 
\begin{equation}
  v_j(\x) = e^{\beta \x_j}\ ,\qquad \nabla v_j(\x) = \beta e_j v_j(\x) \ ,\qquad 1\leq j\leq 3
\end{equation}
are solutions of the above equation. Next we define 
\begin{equation}
 v_4(\x) = x_1 e^{\beta x_2}\ , \qquad \nabla v_4(\x) =   (e_1+x_1\alpha e_2) e^{\beta x_2} \ ,
\end{equation}
which is also easily verified to be a solution of the above equation. Now let $B$ be the $4\times 4$ matrix defined by 
\begin{equation}
B = \left(\begin{matrix}
v_1 & v_2 & v_3 & v_4 \\  \grad v_1 & \grad v_2 & \grad v_3 &  \grad v_4
\end{matrix}\right) \ .
\end{equation}
We find that for $\phi(\x)=e^{\beta(x_1+2x_2+x_3)}$,
\begin{equation}
 {\rm det} B = \phi(\x) {\rm det} \left(\begin{matrix}
  1 & 1 & 1 & x_1 \\ \beta e_1 & \beta e_2 & \beta e_3 & (e_1+x_1\beta e_2)
\end{matrix}\right)=\phi(\x) {\rm det} \left(\begin{matrix}
  1 & 1 & 1 & 0 \\ \beta e_1 & \beta e_2 & \beta e_3 & e_1
\end{matrix}\right)=\beta^2 \phi(\x).
\end{equation}
We see that $\det B$ does not vanish and hence the four solutions so constructed are linearly independent. By continuity, any family of solutions with boundary conditions $f_j$ close to the restrictions of the $v_j$ to $\partial \Omega$ will generate a family of functions $\{v_j\}$ such that $\{(\nabla v_j,v_j)\}_{1\leq j\leq 4}$ forms a basis of ${\mathbb R}^4$ at every point $\x\in{\mathbb R}^3$.
It is important to note that the above solutions do not extend to arbitrary (positive) non-constant coefficients. In such a setting we can use the  techniques of complex geometrical optics and Runge approximation developed, for example, in \cite{bal_2012_a,bal_2013}. 

We now consider solutions of the form $u=e^{\Brho\cdot \x}$, where $\Brho=\Brho_r+i\Brho_i$ is a complex-valued vector in ${\mathbb C}^n$ such that $\Brho_r\cdot\Brho_i=0$ and $\rho_r^2=\rho_i^2$. This implies that $\Brho\cdot\Brho=0$ and hence  $\Delta e^{\Brho\cdot \x}=0$. For $|\Brho|$ fixed, we define four such vectors
\begin{equation}
\Brho_1=|\Brho|(e_1+ie_2),\quad \Brho_2=|\Brho|(e_1-ie_2),\quad \Brho_3=2|\Brho|(e_1+ie_3),\quad \Brho_4=2|\Brho|(e_1-ie_3),
\end{equation}
as well as the four harmonic solutions $u_j=e^{\Brho_j\cdot \x}$ for $1\leq j\leq 4$. These solutions are complex-valued, but we can define the real-valued harmonic functions $v_1=\Re u_1$,  $v_2=\Im u_1$,  $v_3=\Re u_3$,  $v_4=\Im u_3$, so that $u_1=v_1+iv_2$, $u_2=v_1-iv_2$, and so on, and realize that the matrix $B$ is invertible if and only if 
\begin{equation}
\tilde B = \left(\begin{matrix}
u_1 & u_2 & u_3 & u_4 \\  \grad u_1 & \grad u_2 & \grad u_3 &  \grad u_4
\end{matrix}\right) \ .
\end{equation}
is invertible. 
Then we find that
\begin{equation}
\dfrac{{\rm det} B}{u_1u_2u_3u_4} = \det \left(\begin{matrix}
  1 & 1 & 1 & 1 \\ \rho_1 & \rho_2 & \rho_3 & \rho_4
\end{matrix}\right)=|\rho|^4 \det \left(\begin{matrix}
  1 & 1 & 1 & 1 \\ e_1 & -2ie_2 & 2e_1 & -4ie_3
\end{matrix}\right)=-8|\rho|^4.
\end{equation}
Thus we conclude that the family $\{(\nabla v_j,v_j)\}_{1\leq j\leq 4}$  forms a basis of ${\mathbb R}^4$ and that the inverse of the matrix $B$ is bounded uniformly on a bounded domain. It remains to consider the case of non-constant coefficients. The theory developed in \cite{bal_2010,bal_2012_a} shows that we can construct solutions of $-\nabla\cdot D_0(\x)\nabla u+\alpha_0(\x) u=0$ of the form
\begin{equation}
   u_j(\x) = \dfrac{1}{\sqrt{D_0(x)}}e^{\Brho_j\cdot \x} (1+ \psi_j(\x)) \ ,
\end{equation}
with $\rho_j$ as above and $\psi_j$ arbitrarily small  for $|\Brho|$ sufficiently large. Thus choosing $|\Brho|$ sufficiently large, $\tilde B$ is invertible, as is the corresponding matrix based on the solutions of the elliptic equation $v_1=\Re u_1$,  $v_2=\Im u_1$,  $v_3=\Re u_3$,  $v_4=\Im u_3$ (since the coefficients are real-valued). Now, this property remains true if the boundary conditions $\{f_j\}$ are chosen to be sufficiently close to the restrictions of the above solutions $u_j$ on $\partial \Omega$. Such conditions are, however, not explicitly known, but the above theory shows that an open set of such conditions exists.